\documentclass[review,number]{elsarticle}
\usepackage{graphicx}
\usepackage{lineno}
\usepackage{subcaption}
\usepackage{hyperref}
\usepackage{bm}

\usepackage{upgreek}
\usepackage{textcomp}
\usepackage{amssymb}
\usepackage{amsthm}
\usepackage{url}

\usepackage{amsmath}
\usepackage{esint}
\usepackage{lineno}
\usepackage{color}

\usepackage{xcolor}
     \usepackage[left=2.54cm,right=2.54 cm,top=2.54 cm,bottom=2.54 cm]{geometry} 

\begin{document}
\begin{frontmatter}

\def\ps@pprintTitle{%
 \let\@oddhead\@empty
 \let\@evenhead\@empty
    \let\@evenfoot\@oddfoot}

\title{Attenuation of electromagnetic radiation in Nuclear Track Detectors}

\author[my0address]{R. Bhattacharyya\footnote{Corresponding author. E-mail address: rupamoy@gmail.com}}
\author[my2address]{A. Maulik\footnote{Present address: Istituto Nazionale di Fisica Nucleare, Sezione di Bologna, Bologna 40127, Italy}}
\author[my0address,mysecondary2address]{R. P. Adak}
\author[my0address]{S. Roy}
\author[mysecondaryaddress]{T. S. Bhattacharya}
\author[my0address,mysecondaryaddress]{S. Biswas}
\author[my0address,mysecondaryaddress]{S. Das}
\author[my0address]{S. Dey}
\author[my0address,mysecondaryaddress]{S. K. Ghosh}
\author[my3address]{K. Palodhi}
\author[my0address,mysecondaryaddress]{S. Raha}
\author[mysecondaryaddress]{A. Singha}
\author[my0address]{D. Syam}

\address[my0address]{Department of Physics $\&$ Centre for Astroparticle Physics and Space Science, Bose Institute, Kolkata 700091, India}
\address[my2address]{Physics Department, University of Alberta, Edmonton T6G 2E1, Canada}
\address[mysecondary2address]{Department of Physics, Taki Government College, Taki 743429, India}
\address[my3address]{Department of Applied Optics and Photonics, University of Calcutta, Kolkata 700106, India}
\address[mysecondaryaddress]{Department of Physics, Bose Institute, Kolkata 700009, India}

\begin{abstract}
A systematic study of the attenuation of electromagnetic radiation in Nuclear Track Detectors (NTDs) is carried out. The attenuation of gamma-ray, X-ray, UV, visible, and infrared radiation in NTDs are investigated using NaI(Tl) detector, Gas Electron Multiplier (GEM) detector, UV-Vis spectrophotometer, and FTIR spectrophotometer respectively. 
The values of some important parameters (e.g., optical depth, attenuation coefficient, etc.) of three commercially available NTDs (PET, Makrofol$^{\circledR}$ and CR-39$^{\circledR}$), at the relevant region of the electromagnetic spectrum, is determined. The details of the experimental techniques and the results are also presented in this paper.
\end{abstract}

\begin{keyword}
Nuclear track detector\sep Electromagnetic radiation \sep Radiation detector\sep  Attenuation coefficient \sep UV blocker
\end{keyword}
\end{frontmatter}

\section{Introduction}
Nuclear Track Detectors (NTDs) have often been the detectors of choice for detecting highly ionizing rare particles (strangelet, magnetic monopole) in cosmic rays~\cite{Cecchini:2008su,Cecchini:2009br} or in accelerators~\cite{Acharya:2014nyr,article}. Some of the
most important advantages of NTDs are low cost, ease of use, and the existence of natural thresholds of detection~\cite{BHATTACHARYYA201663}. NTDs offer excellent charge($\lesssim1$ $e$, where $e$ is the electronic charge) and position
resolution ($\sim 1\upmu$m)~\cite{BALESTRA2007254,Dey:2014tza}. In addition, being passive in the nature of action (as they do not require electrical power during their operation), NTDs are often the best choice for the experiments which require detector arrays to be set up in remote locations for cosmic ray observation~\cite{Balestra:2008ps,Cecchini:2008su}. Previous studies show that the responses of the NTDs (e.g., bulk etch-rate, sensitivity, etc.) are largely affected both in open-air and in accelerator by the exposure due to ambient electromagnetic radiations~\cite{SHARMA1991385,JAIN2020244,1475-7516-2017-04-035}. 
To investigate this particular issue, we pursued a systematic study on the attenuation of electromagnetic radiation over a wide range of photon energies for three commercially available NTDs, namely - PET (Desmat, India), Makrofol$^{\circledR}$ (Covestro AG, Germany) and CR-39$^{\circledR}$ (TASL, England). The attenuation of gamma-ray, X-ray, UV-visible, and infrared radion in NTDs is studied using NaI(Tl) detector, Gas Electron Multiplier (GEM) detector, UV-Vis spectrophotometer, and FTIR spectrophotometer respectively. The values of some important parameters (e.g., optical depth, attenuation coefficient, etc.) of different NTDs at the relevant region of the electromagnetic spectrum is determined.  
\par 
Electromagnetic radiation interacts with matter through three processes: 
(i) photoelectric effect, (ii) Compton scattering, and (iii) pair production. While passing through the transparent material, the number of transmitted photons get reduced by scattering and absorption due to the processes mentioned above. 
It can be shown that the intensity of the photons $I_0$, after passing through a distance $x$ inside the absorber reduces to
\begin{equation}
 I(x)=I_0e^{-\mu x}
\end{equation}
where $\mu$, the \emph{attenuation coefficient}, is a characteristic of the 
absorbing material and the energy of the photon~\cite{Leo:1987kd}.
\par
The details of the measurement techniques, experimental setup, and the results of the study of the attenuation of electromagnetic radiation (gamma-ray, X-ray, UV ray, visible ray, and infrared ray in particular)
while passing through the NTDs, are presented in the next few sections.

\section{Experimental arrangement}

Nuclear Track Detectors (NTDs) are optically transparent, dielectric solids comprising of long polymer chains. The chemical composition~\cite{Durrani:1987zv} and thickness of the different plastic NTDs used in these experiments are given in Table~1.

\begin{table}[h]
\centering
\begin{tabular}{|c|c|c|c|c|}
\hline 
\hline
\bf NTD material	& \bf Chemical composition				& \bf Thickness	\boldmath ($\upmu$m)			\\
\hline
 PET & (C$_{10}$H$_8$O$_4$)$_\textnormal{n}$& 90\\
 (Poly-ethylene terephthalate)  	& 					& 		\\
 \hline
 Makrofol$^{\circledR}$ 			& (C$_{16}$H$_{14}$O$_3$)$_\textnormal{n}$		& 480		\\
 (Bisphenol-A polycarbonate)  	& 					& 		\\
\hline
 CR-39$^{\circledR}$			& (C$_{12}$H$_{18}$O$_7$)$_\textnormal{n}$		& 630		\\
 (Poly-allyl diglycol carbonate)  	& 					& 		\\
\hline 
\hline
\end{tabular}
\caption{Name, chemical composition and thickness of different NTDs used in this experiment.}
\label{NTDdetails}
\end{table}
The detailed experimental procedures in different electromagnetic regions are described below.
\subsection{\bf{Gamma ray region using NaI(Tl) scintillation detector}}
\begin{figure} 
 \centering
     \begin{subfigure}[b]{0.495\textwidth}
         \centering
         \includegraphics[width=\textwidth, height=0.8\textwidth]{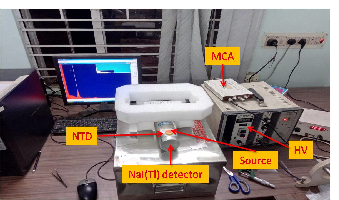}
         \caption{}
     \end{subfigure}
          \begin{subfigure}[b]{0.495\textwidth}
         \centering
         \includegraphics[width=\textwidth, height=0.8\textwidth]{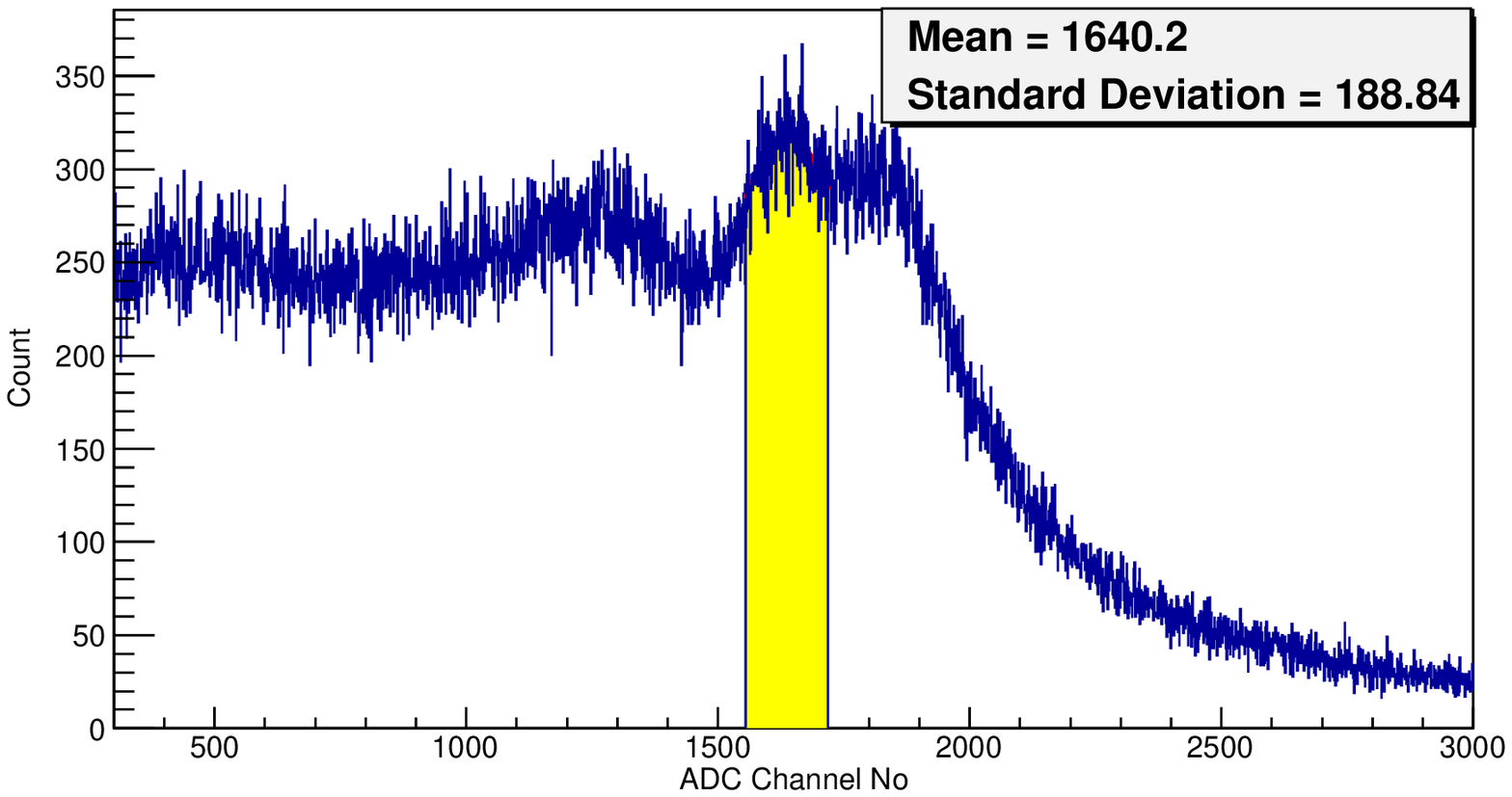}
               \caption{}
     \end{subfigure}
\caption{(a) Experimental set up for the study of attenuation of $\gamma$ rays by NTDs using  NaI(Tl) inorganic scintillator detector. (b) Energy spectrum of $^{60}$Co radioactive source by NaI(Tl) detector.}
\label{NaI}
\end{figure}

A NaI(Tl) scintillator detector~\cite{Knoll:1300754} coupled with a Photo Multiplier Tube (PMT) is used to detect the gamma radiation from a $^{60}$Co source.
A high voltage of 600 V is applied to the PMT. 
The signal from the PMT is fed to a charge sensitive pre-amplifier which is integrated with the base of the PMT. The pre-amplifier signal is fed to a  spectroscopy amplifier (ORTEC-671) with a coarse gain of 500. The amplified signal is then connected to a PC controlled Multi Channel Analyser (MCA) (Ortec EASY-MCA-8k) to obtain the energy spectrum.
The attenuation of $\gamma$ rays is studied by placing the NTDs in-between 
the source and the detector, as shown in Fig.~\ref{NaI}(a). The energy spectrum in the given range ($300-3000$) of the ADC channel number is plotted in Fig.~\ref{NaI}(b). The first peak is due to 
$^{60}$Co source of energy 1.173 MeV and is fitted with a Gaussian function.

\begin{figure}[h]
 \centering
     \begin{subfigure}[b]{0.495\textwidth}
         \centering
         \includegraphics[width=\textwidth, height=0.8\textwidth]{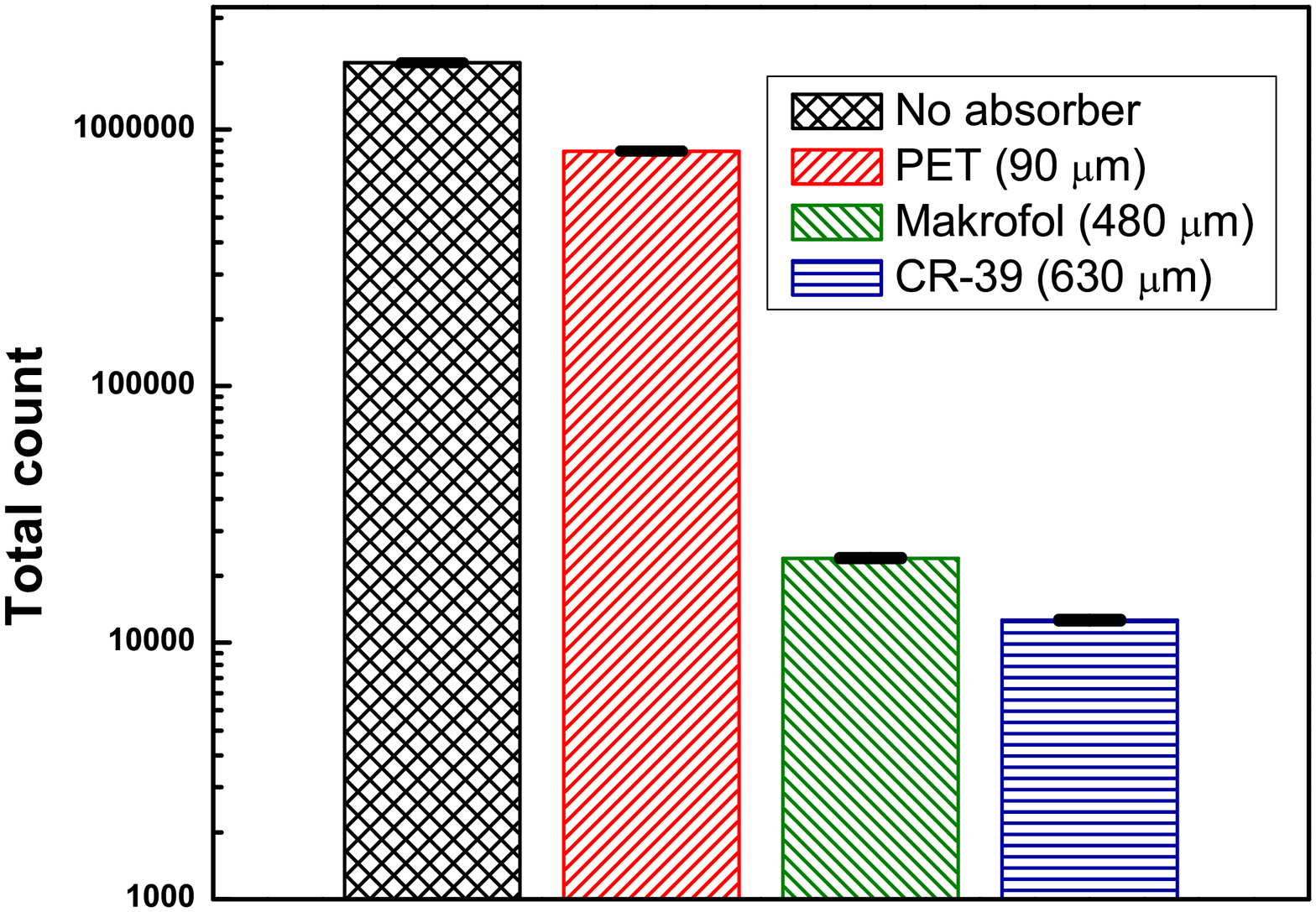}
         \caption{}
     \end{subfigure}
          \begin{subfigure}[b]{0.495\textwidth}
         \centering
         \includegraphics[width=\textwidth, height=0.8\textwidth]{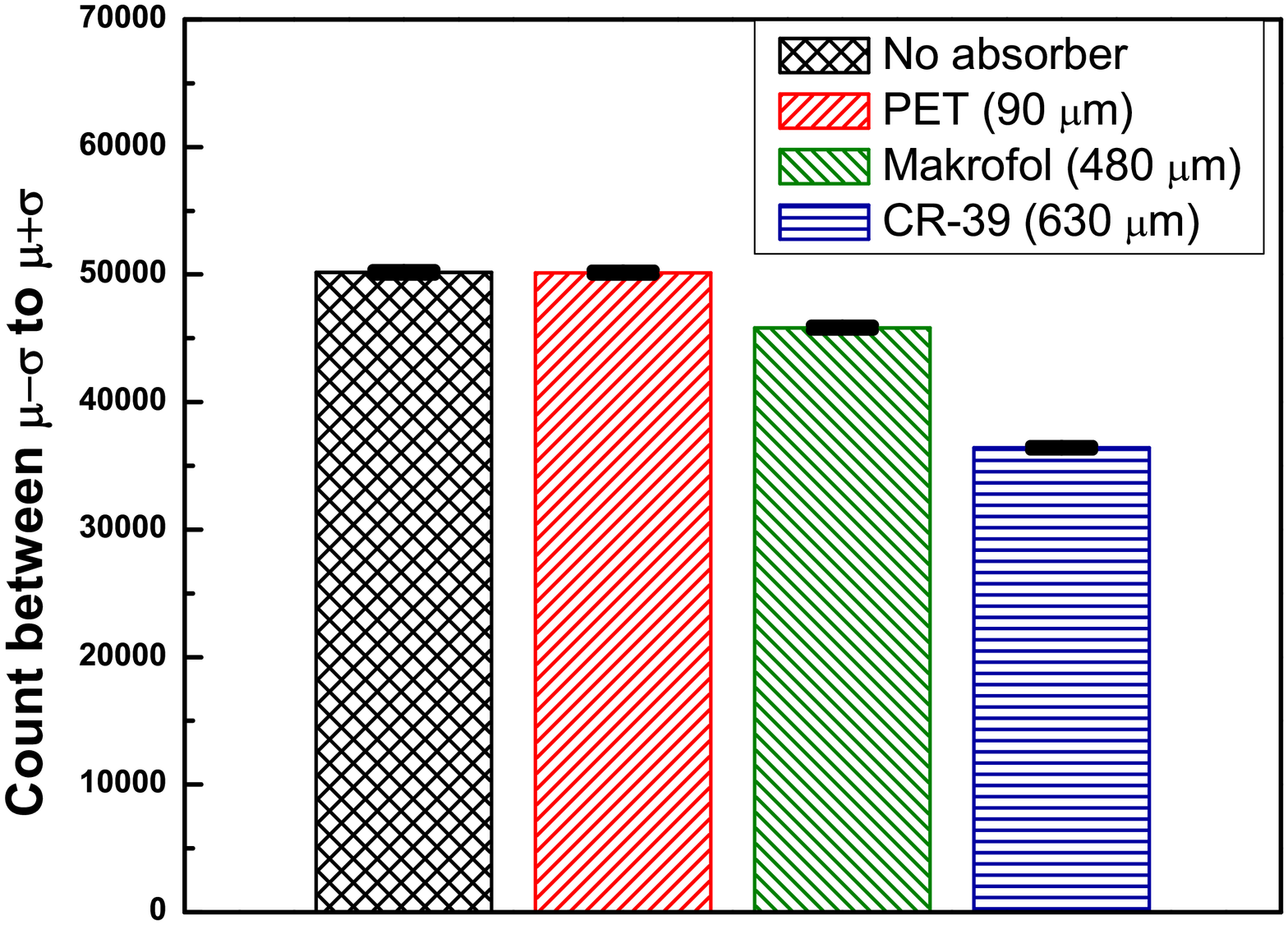}
         \caption{}
     \end{subfigure}
\caption{(a) Attenuation of X-ray photons by NTDs. 
    (b) Attenuation of $\gamma$ ray photons by NTDs.}
\label{GEM_NaI_NTD}
\end{figure}

Total number of counts within one standard deviation ($\pm1\sigma$) from the mean ($\mu$); i.e., between 
ADC channel number 1555 to 1718 is plotted in a histogram (Fig.~\ref{GEM_NaI_NTD}(b)). The NTDs are placed one by one before the scintillation counter. From Fig.~\ref{GEM_NaI_NTD}(b) the attenuation coefficient of 
different NTDs in the energy range $1.2\pm0.8$ MeV can be calculated and
the values are presented in Table~\ref{AttenuationCoefficient}.


\subsection{\bf{X-ray region using Gas Electron Multiplier (GEM) detector}}

\begin{figure}[h]
 \centering
     \begin{subfigure}[b]{0.495\textwidth}
         \centering
         \includegraphics[width=\textwidth, height=0.8\textwidth]{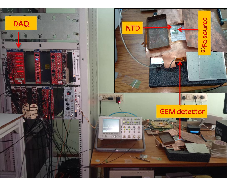}
       \caption{}
     \end{subfigure}
          \begin{subfigure}[b]{0.495\textwidth}
         \centering
         \includegraphics[width=\textwidth, height=0.8\textwidth]{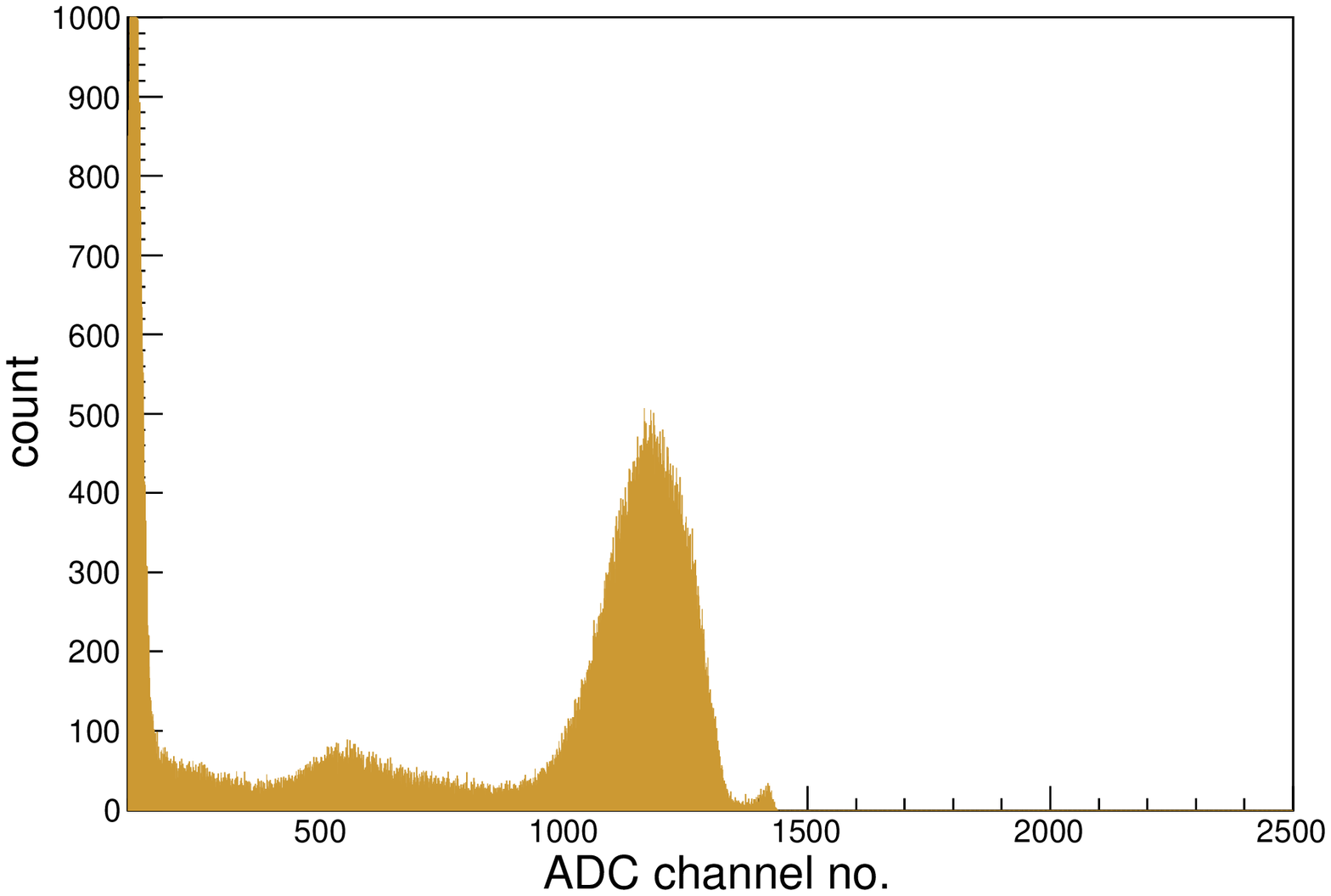}
      \caption{}
     \end{subfigure}
\caption{(a) Experimental set up for the study of attenuation of X-rays by NTDs using GEM detector. (b) Energy spectrum of $^{55}$Fe radioactive source by the GEM detector.}
\label{GEM} 
\end{figure}

In this experimental setup,
X-ray radiation is detected by a Gas Electron Multiplier (GEM) detector~\cite{SAULI1997531,1748-0221-11-10-T10001,doi:10.1080/10739149.2020.1742155}. 
Triple GEM detector is a gaseous ionization detector; it consists of three GEM foils cascaded inside a chamber filled with gas. 
A pre-mixed gas mixture of Ar:CO$_2$ in 70:30 volume ratio is used at a flow rate of 3.5 l$/$h.
Each GEM foil (10 cm$\times$10 cm) 
is made up of a polymer (Kapton)
foil of thickness $50~\upmu$m sandwiched between two copper foils of thickness $5~\upmu$m.
Relatively low voltages ($\sim~400$ V) are applied between the copper planes of each foil.
There are a large number of holes 
($\sim80$ mm$^{-2}$) of diameter $70~\upmu$m etched on the GEM foil at a pitch of $140~\upmu$m. 
Thus a high electric field ($\sim80$ kV cm$^{-1}$) is created inside the holes. A negative high voltage of $-4175$ V is distributed across the GEM foils and the electrodes using a passive resistive chain.
 Ionizing radiation, on passing through the gas, creates electron-ion pairs. An electron, while moving along the large electric field inside those holes, initiates an avalanche of electrons. They are then collected from the readout anode pad placed after the third GEM foil. The experimental arrangement is shown in Fig.~\ref{GEM}(a).
X-rays from $^{55}$Fe source get partially attenuated by the NTD placed in-between the source and the detector.
The signal from the GEM detector is fed to a charge sensitive pre-amplifier (VV50-2) \cite{vv02}. The analog signal is then fed to 
a four output linear Fan-in-Fan-out
(linear FIFO) module in a NIM based data acquisition system. One of the outputs is connected to a MCA to obtain the energy spectrum. One such spectrum of $^{55}$Fe on GEM detector is shown in Fig.~\ref{GEM}(b) where the noise peak (below channel number 150), argon escape peak (channel number 570, energy 2.9 keV), and main 5.9 keV X-ray peak (channel number 1190) from $^{55}$Fe source are clearly visible.
Another output from the linear FIFO is fed to a Single Channel Analyzer (SCA), the threshold of which is set at 0.1 V to reject the noise. 
The total count of the incident particles is measured by a NIM scaler, which is connected to the SCA via a TTL-NIM adapter. The findings of this study are presented in the results and discussions section.

\subsection{\bf{Visible to ultraviolet region using UV-Vis spectrophotometer}}
\begin{figure}[h]
\centering
     \begin{subfigure}[b]{0.495\textwidth}
         \centering
         \includegraphics[width=\textwidth, height=0.8\textwidth]{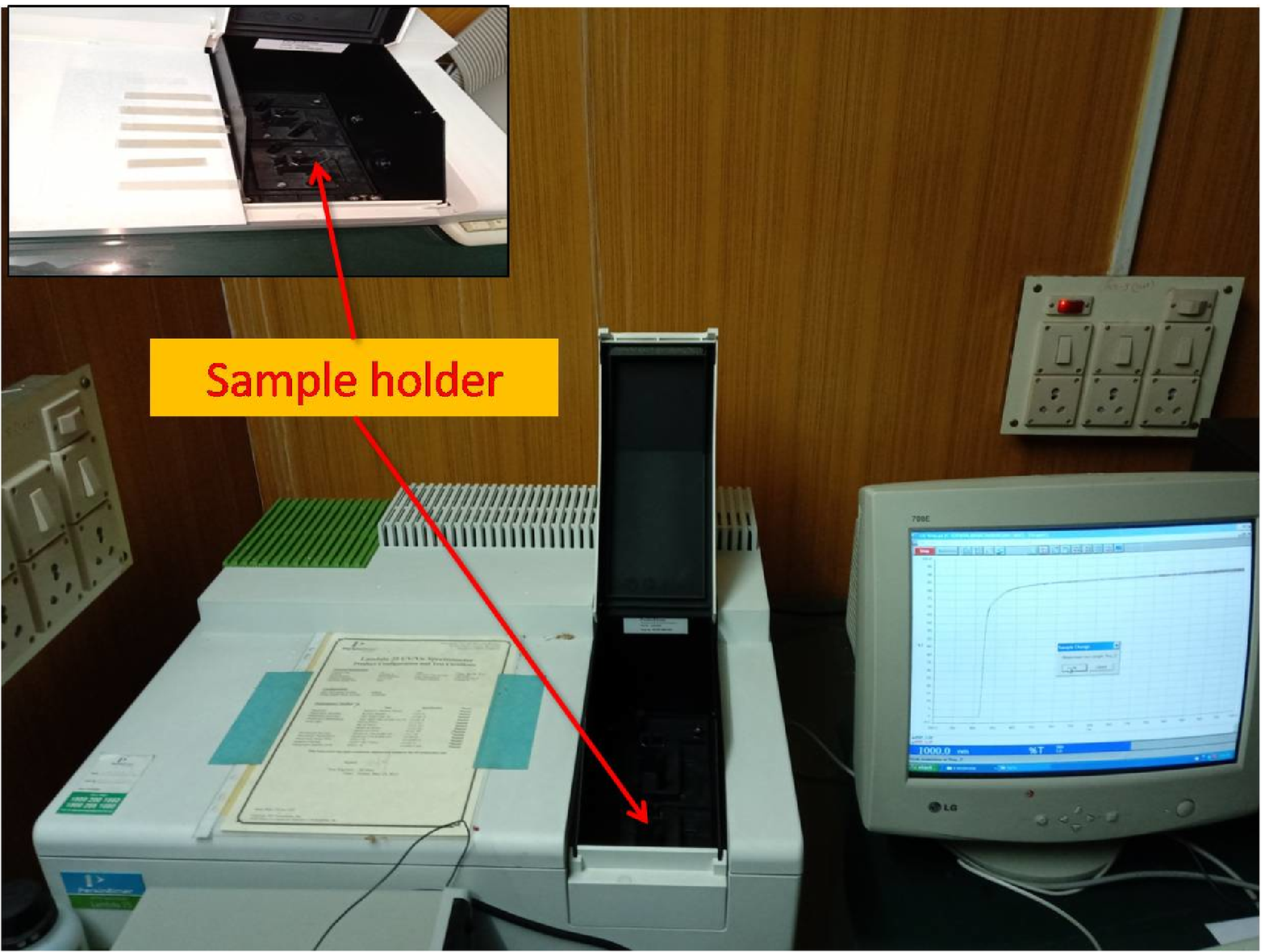}
         \caption{}
     \end{subfigure}
          \begin{subfigure}[b]{0.495\textwidth}
         \centering
         \includegraphics[width=\textwidth, height=0.8\textwidth]{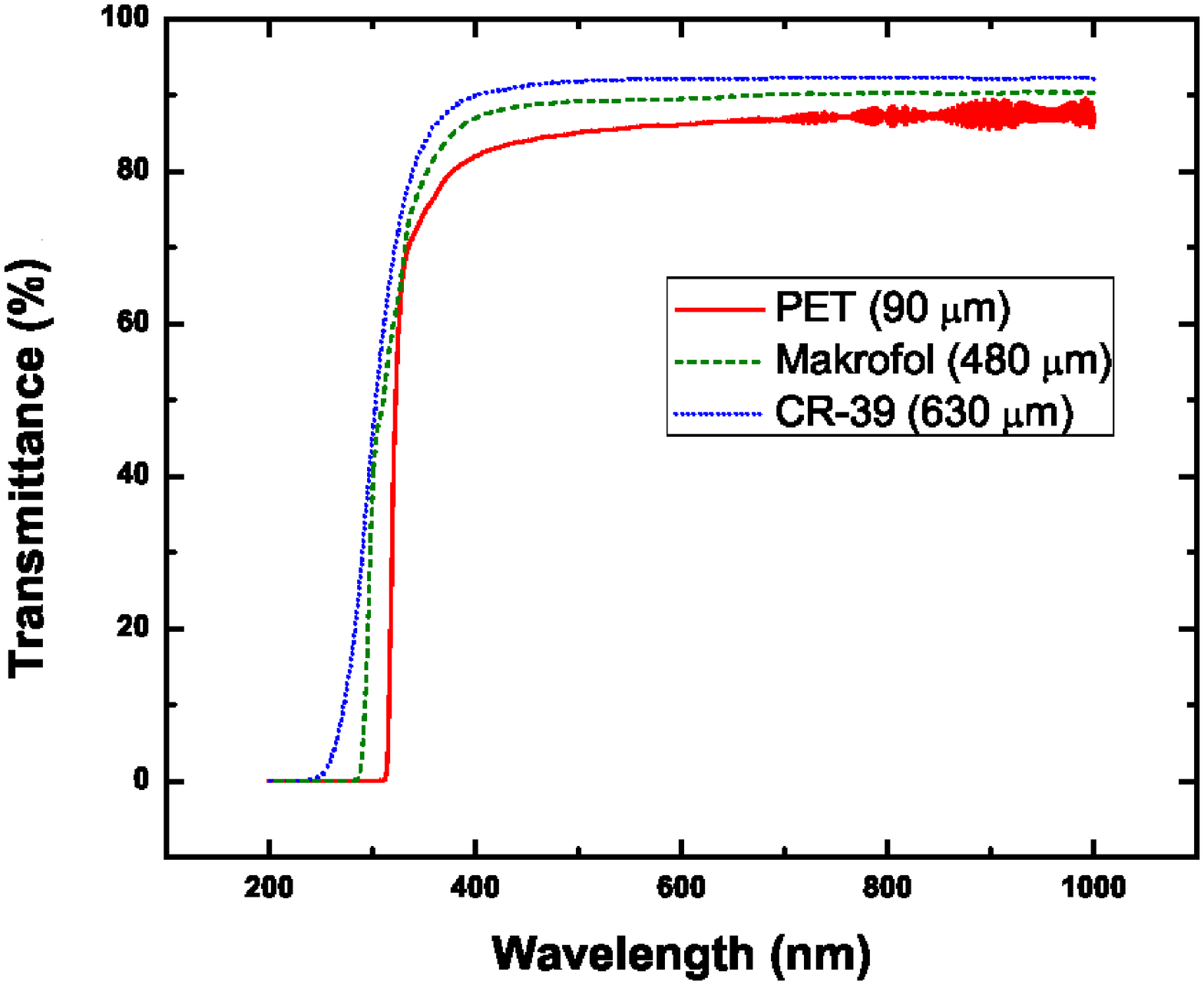}
         \caption{}
     \end{subfigure}	
\caption{(a) Instrument for the study of attenuation of UV and visible radiation in NTDs. 
    (b) Attenuation of photons in ultra-violet and visible regions by NTDs. }
\label{UVVis} 
\end{figure}

To check the transmittance of different NTD samples (Table~\ref{NTDdetails}) in Ultra-Violet (UV) and visible radiation, they are scanned by a Perkin-Elmer Lambda 25 UV/Vis spectrometer (Fig.~\ref{UVVis}(a)), 
covering wavelength ranging from 200 nm to 1000 nm with a resolution of 1 nm. Fig.~\ref{UVVis}(b) shows the transmission profile of different NTDs in the visible and UV region of electromagnetic radiation.

\begin{figure}[h]
\centering 
     \begin{subfigure}[b]{0.495\textwidth}
         \centering
         \includegraphics[width=\textwidth, height=0.8\textwidth]{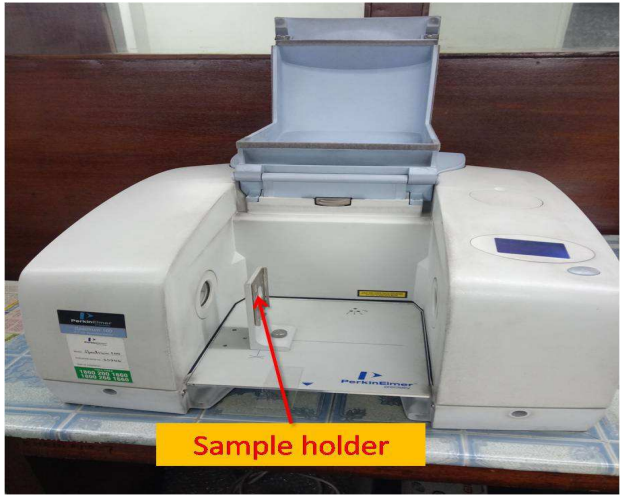}
       \caption{}
     \end{subfigure}
          \begin{subfigure}[b]{0.495\textwidth}
         \centering
         \includegraphics[width=\textwidth, height=0.8\textwidth]{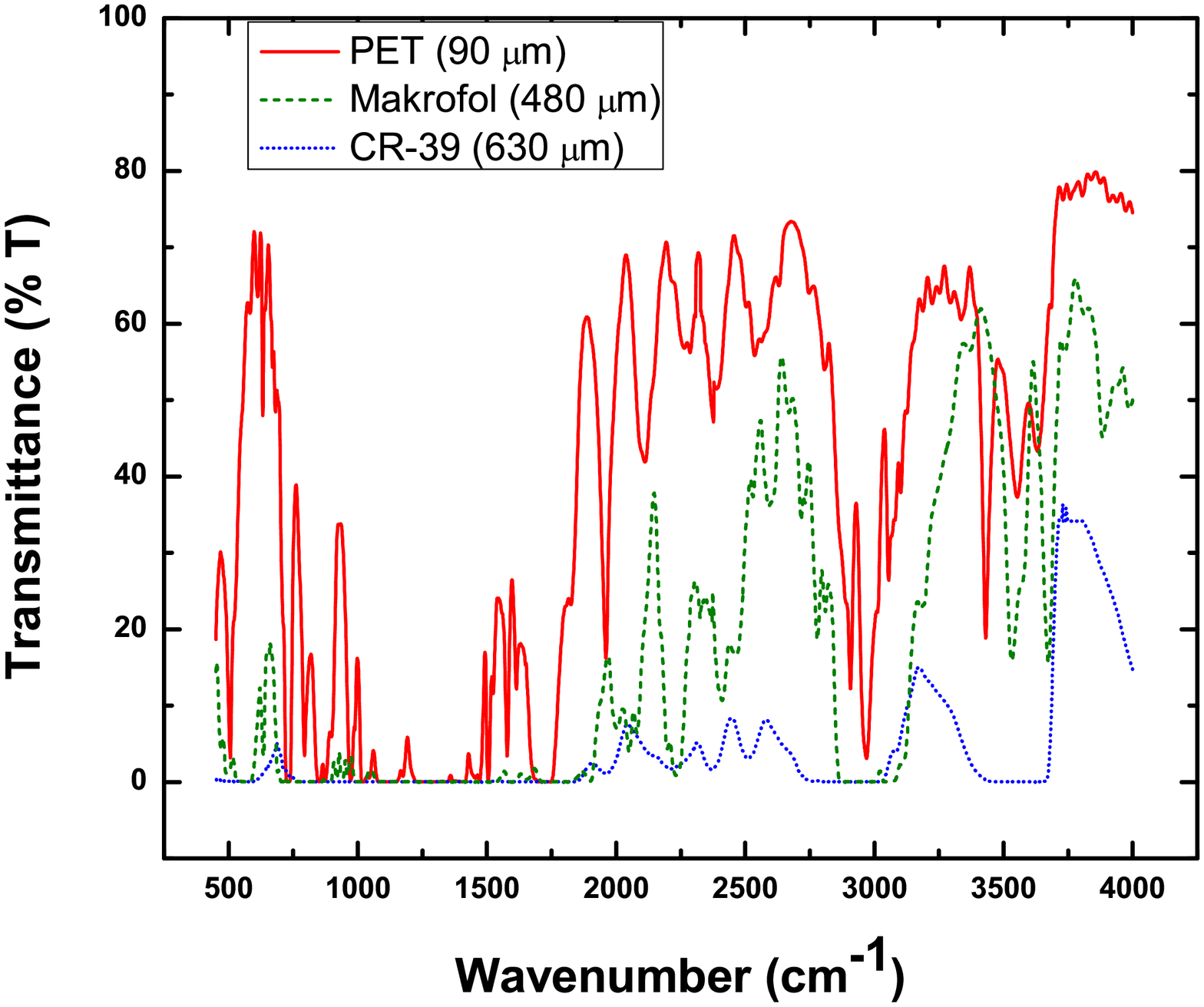}
         \caption{}
     \end{subfigure}
  
\caption{(a) Instrument for the study of attenuation of infrared radiation in NTDs. 
    (b) Attenuation of the number of photons in the infrared region by NTDs.}
\label{FTIR} 
\end{figure}

\subsection{\bf{Infrared region using FTIR spectrophotometer}}
NTD samples (PET, Makrofol$^{\circledR}$ and CR-39$^{\circledR}$) are analyzed by Fourier Transform Infra-Red (FTIR) spectroscopy~\cite{1994fundamentals}(wavenumber ranging from 450 to 4000 cm$^{-1}$) using Perkin-Elmer Spectrum 100 FT-IR spectrometer (Fig.~\ref{FTIR}(a)).
The infrared transmission spectrum of the samples is shown in Fig.~\ref{FTIR}(b).

\section{Results and discussions}
From Fig.~\ref{GEM_NaI_NTD}(a) the attenuation coefficient of different NTDs in the energy range $1-7$ keV can be calculated and the values are presented in Table~\ref{AttenuationCoefficient}. It is clear that the attenuation coefficient for all three types of NTDs have nearly same values in the X-ray region. Whereas, in the gamma ray region, the value of the attenuation coefficient for PET is smaller by two orders of magnitude than the other two NTDs (CR-39$^{\circledR}$ and Makrofol$^{\circledR}$). 
\begin{table}[h]
\centering
\begin{tabular}{|c|c|c|c|c|}
\hline 
\hline
\bf Name	&  \multicolumn{2}{|c|}{\bf \boldmath Attenuation coefficient $\mu$ (m$^{-1}$)}    \\ 
		& \bf \boldmath $\gamma$ ray region ($1.2\pm0.8$ MeV)  & \bf X-ray region (1-7 keV)  \\
\hline
 PET 			& $5.3\times 10^{-12}$ 	& $8.8\times 10^{-9}$			\\
 \hline
 Makrofol$^{\circledR}$ 			& $1.9\times 10^{-10}$ 	& $9.3\times 10^{-9}$ 		\\
\hline
 CR-39$^{\circledR}$			& $5.1\times 10^{-10}$ & $7.9\times 10^{-9}$ 		\\
\hline 
\hline
\end{tabular}
\caption{Attenuation coefficient of different NTDs in the gamma ray and X-ray region respectively. Experimental uncertainties in the measurements are $<~1\%$ \cite{1748-0221-11-10-T10001,Roy:2019jwy}.}
\label{AttenuationCoefficient}
\end{table}
\par 

Fig.~\ref{UVVis}(b) shows that the transmittance near the visible region of electromagnetic radiation is more than $85\%$ for all NTDs, which is helpful to use them in transmitted light optical microscopy. Optical depths ($\tau$) corresponding to the wavelength 560 nm (yellow light) are calculated using the relation
\begin{equation}
\tau=ln\frac{\phi^i}{\phi^t}
\end{equation}
where $\phi^i$ and $\phi^t$ are the radiant flux received and transmitted by the material at that wavelength. For PET, Makrofol$^{\circledR}$ and CR-39$^{\circledR}$ the values of $\tau$ (at wavelength 560 nm) are given in Table~\ref{PhysicalPara}. 
\begin{table}[h]
\centering
\begin{tabular}{|c|c|c|c|}
\hline 
\hline
\bf NTD  & \bf Optical & \bf Wavelength & \bf Direct band-gap\\
\bf name & \bf depth ($\tau$) & \bf cut-off (nm)    & \bf energy $E_g$ (eV)\\
\hline
 PET 			& $ 1.03$ 	& $310$ & $3.82$			\\
 \hline
 Makrofol$^{\circledR}$ 	& $1.02$ 	& $280$ & $3.94$ 		\\
\hline
 CR-39$^{\circledR}$	& $1.02$ & $230$ & $3.98$		\\
\hline 
\hline
\end{tabular}
\caption{Physical parameters of different NTDs corresponding to ultraviolet and visible region of electromagnetic wavelength.}
\label{PhysicalPara}
\end{table}
\par
\begin{figure}[h]
 \centering
 \includegraphics[width=1.0\textwidth]{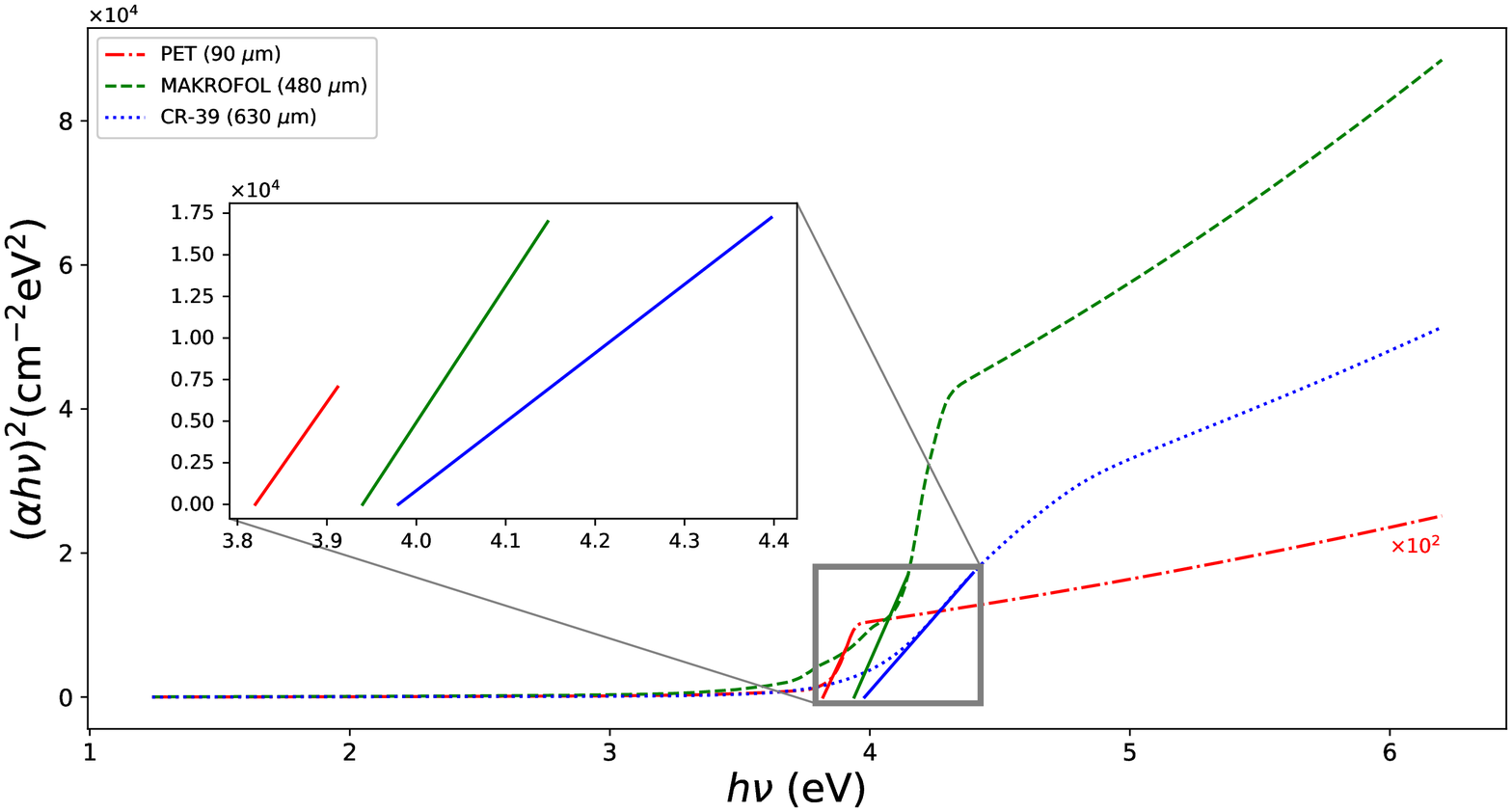}
 \caption{Plot of $(\alpha h \nu)^2$ vs. photon energy ($h\nu$) for different NTDs. Extrapolation the linear part of the curve is shown in the inset.}
 \label{AlphavsE}
\end{figure}

 
The optical energy gap ($E_g$) is related to the absorption coefficient ($\alpha$) through the relation~\cite{Mott},
\begin{equation}
\alpha (h \nu) = \frac{B(h\nu-E_g)^{1/2}}{h\nu}
\end{equation}
where $\alpha$ is calculated from $\alpha=2.303\times A/ t$ ($A$ and $t$ are the absorbance and thickness of the material respectively); $B$ is a constant. By plotting $(\alpha h \nu)^2$ as a function of photon energy ($h\nu$) [Fig.~\ref{AlphavsE}] and extrapolating the linear part of the curve to $h\nu$-axis at zero absorption, optical energy gap for different NTDs are computed~(\cite{TAYEL2015219}) and the corresponding values are given in Table~\ref{PhysicalPara}.
Fig.~\ref{UVVis}(b) shows that the value of transmittance is negligibly small ($\simeq0$) below the wavelengths 310 nm, 280 nm, and 230 nm for PET, Makrofol$^{\circledR}$ and CR-39$^{\circledR}$ respectively (Table~\ref{PhysicalPara}). This shows that PET film blocks UV radiation more effectively than the other two.
Thus, while using stacks of NTDs, keeping PET film in the top layer during open-air exposures,
the effect of UV rays on altering 
the etch-rate may be prevented in the lower layers of NTDs~\cite{1475-7516-2017-04-035}. The fact that PET can block UV radiation can have other practical uses as well. For example, UV radiation is effective in disinfecting SARS-CoV-2 surface contamination \cite{Kitagawa2020}. But prolonged exposure to UV can pose a health hazard to humans. So PET films can offer a cost-effective way to shield humans from the harmful effects of UV.
\par

Results from the FTIR spectroscopy (Fig.~\ref{FTIR}(b)) is useful in the identification of polymers (i.e., their chemical composition), by identifying the peaks due to different functional groups~\cite{Yamauchi_2012}. However, as we are interested here in the attenuation of electromagnetic radiation in NTDs, these results are not considered for further study.

\section{Conclusion}

We have estimated some of the relevant physical quantities (like attenuation coefficient, optical depth) of previously unexposed NTDs over a wide range of the electromagnetic spectrum. These results will serve as a reference in the study of the response (i.e., bulk etch-rate, sensitivity) of NTDs exposed to electromagnetic radiation over a long period of time. In this work, it is also shown that a single PET film of thickness $90~\upmu$m can block the UV radiation almost completely. So, when a stack of PET films is given open-air exposure, the topmost PET film can prevent the change of etch-rate due to UV related damage in all subsequent layers. The ability of PET to block UV rays can find applications in other fields of research and industry.

\section*{Acknowledgments} 
RB wishes to thank Dr. Rajarshi Ray, Bose Institute for his valuable suggestions and useful discussions. This work is partially funded by IRHPA (Intensification of Research in High Priority Areas) Project (IR/S2/PF-01/2011 dated 26.06.2012) of the Science and Engineering Research Council (SERC), DST, Government of India, New Delhi.

\bibliography{mybibfile.bib}

\end{document}